\documentclass[preprint,onecolumn,10pt]{revtex4-1}
\usepackage{graphicx}
\usepackage{color}
\usepackage{amsmath}
\usepackage{multirow}

\begin{document}

\title{First-principles study of electronic structure of $Bi_2Sr_2Ca_2Cu_3O_{10}$}
\author{J. A. Camargo-Mart\'inez}
\affiliation{Departamento de F\'isica, CINVESTAV-IPN, Av. IPN 2508, 07360 M\'exico}
\author{Diego Espitia}
\affiliation{Departamento de F\'isica, CINVESTAV-IPN, Av. IPN 2508, 07360 M\'exico}
\author{R. Baquero}
\affiliation{Departamento de F\'isica, CINVESTAV-IPN, Av. IPN 2508, 07360 M\'exico}

\begin{abstract}

We present for the first time the band structure calculation of $Bi_2Sr_2Ca_2Cu_3O_{10}$ compound in the tetragonal structure (space group $I4/mmm$). We used the Local Density 
Approximation (LDA) as in the Wien2k code. We analyze in detail the band structure and the Fermi surface (FS). Our results are in very good agreement with recent experiments. The FS 
shows the feature known as the Bi-O pocket problem which we associate with the interaction of the O3 atoms with the Cu2-O2 and Bi-O4 planes. Ceramic $Bi_2Sr_2Ca_2Cu_3O_{10}$ stabilized 
with Pb has been reported as a superconductor with $T_c \sim 100$. CdS microparticles were embedded into the ceramic $Bi_2Sr_2Ca_2Cu_3O_{10}$. The composite did show a 
superconducting phase transition at a lower $T_c \sim 70 K$. At even lower temperatures re-entrant behavior was observed. The sample regain the superconducting state at
$\sim 47 K$ [arXiv:1101.0277 [cond-mat.supr-con]]. This effect is not observe in the ceramic alone. This calculation is useful per ser and also can contribute to a better 
understanding of this particular re-entrant behavior. 

\end{abstract}
\date{\today}
\pacs{ 74.72.Hs ; 71.20.-b  ; 71.18.+y ; 73.20.At}
\keywords{Bi-2223, electronic structure, Band structure, Fermi surface. }


\maketitle

\section{Introduction}

Bismuth cuprates are high-temperature superconductors (HTSC) (except Bi-2201 with $\sim 2 K$) with the general formula $Bi_2Sr_2Ca_{n-1}Cu_nO_{y}$. They are normally referred to 
by the number of CuO$_2$ planes per unit cell, as Bi-2201, Bi-2212, Bi-2223 and Bi-2234 ($n=1, 2$,  $3$ and $4$  respectively). Resistivity, susceptibility and magnetization 
experiments performed on Bi-2212 and Bi-2223 show a transition to the superconducting state at $\sim85 K$~\cite{1,2} and $\sim110 K$~\cite{A1,A2} respectively. 

The electronic properties of Bi-2212 have been extensively studied both theoretically and experimentally~\cite{bo,12b,13b,14b,15b,16b}. In this work we present a detailed study 
of the electronic band structure, the density of states and the Fermi surface for Bi-2223. To our knowledge a theoretical study of this compound has not been yet reported in the 
literature in spite of the fact that the Bi-2223 compound is one of the most suitable HTSC materials for applications~\cite{A,B,C}. Neutron and X-ray diffraction experiments 
suggest that Bi-2223 presents several orthogonal structures with spacial groups $Amaa$, $A2aa$ and $Fmmm$~\cite{D,E,F} and a tetragonal structure with spatial group 
$I4/mmm$~\cite{G,H}. This structure seems to be more stable when doped with Pb~\cite{F,G,H,I,J}. Recently experimental studies of the electronic structure of Bi-2223 by 
angle-resolved photo-emission spectroscopy (ARPES) have been reported~\cite{K,L,M,Ma}. In particular,  Ideta et al.~\cite{M} report a band splitting at the Fermi surface 
in the  nodal direction associated with the outer and inner CuO$_2$ planes (OP and IP). We will comment on this point below.

Further, an interesting behavior was observed in a ceramic composite of Bi-2223 embedded with microscopic CdS particles. The ceramic Bi2223 alone becomes superconducting at 
$\simeq100 K$ and stays superconducting below this temperature. The composite becomes superconducting at a lower  $T_c\simeq75 K$ and it shows a reentrant behavior~\cite{N,O}, 
i.e., by decreasing the temperature below $T_c$ the composite returns to the normal state measured by resistivity experiments. Lowering further the temperature the superconducting 
state reappears. In some systems the normal state is maintained while in others it returns to the superconducting state at a further lower temperature. This phenomenon has been 
observed since long ago~\cite{cava}. The disappearance of the superconducting state is, in general, attributed to the formation of a magnetic sub-lattice with a Neel Temperature below the 
superconducting critical temperature. The re-entrance of the superconducting state is not yet fully understood. The actual composite has the 
nominal formula $CdS/Bi_2Sr_2Ca_2Cu_3O_{10}$. E. D\'iaz-Vald\'es et al.~\cite{O} reported re-entrant superconductivity in this composite at $T_c\simeq47 K$. Our calculation shows no
magnetic moment in crystalline $Bi_2Sr_2Ca_2Cu_3O_{10}$. We will analyze this case in a further publication.

\section{Method of Calculation}
The electronic properties for bct $Bi_2Sr_2Ca_2Cu_3O_{10}$ was determined with the full-potential linearized augmented plane wave method plus local orbital 
(FLAPW+lo)~\cite{11a} within the local density approximation (LDA) using the wien2k code~\cite{12a}. The core states are treated fully relativistically, while for the valence 
states the scalar relativistic approximation is used. We used a plane-wave cutoff of $R_{mt}K_{Max.}$= 8.0 and for the wave function expansion inside the atomic spheres, a maximum 
value of the angular momentum of $l_{max}$= 12 with $G_{max}$= 25.  We choose a $17\times17\times17$ k-space grid which contains 405 points within the irreducible wedge of 
the Brillouin zone. The muffin-tin sphere radii $R_{mt}$ (in atomic units) are chosen as 2.3 for Bi, 2.0 for Sr, 1.9 for both Ca and Cu, and 1.5 for O.

\section{The Bi-2223 crystal structure}
\begin{figure}[hbt!]
\includegraphics[width=0.3\textwidth]{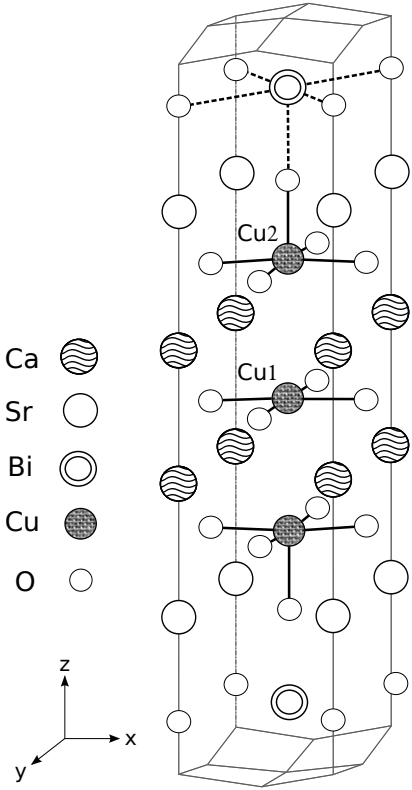} 
\caption{Primitive cell for body-centered tetragonal $Bi_2Sr_2Ca_2Cu_3O_{10}$ O1, O2, O3 and O4 denote oxygens in the Cu1, Cu2, Sr and Bi planes, respectively.}
\label{Fig.a}
\end{figure}

In this work, we study the Bi-2223 compound with body center tetragonal structure (bct) and space group I4/mmm ($D_{4h}^{17}$). The structure consists of three Cu-O planes, 
one Cu1-O1 plane between two Cu2-O2 planes, with Ca atoms between them. Each Cu2-O2 plane is followed by a Sr-O3 and Bi-O4 planes in that order (see Fig.~\ref{Fig.a}). The 
internal parameters were taken  from Ref.~\cite{G}.  

Starting from the experimental parameters, we optimized the {\em c/a} ratio and relaxed the internal coordinates of the structure. In Table~\ref{tabla1} we compare these 
results with the experimental values ​​reported. The optimized {\em c/a} ratio is 1.56$\%$ smaller than the experimental one. In the relaxed structure the Cu1-O1 and Cu2-O2 
planes approach each other by $\sim$0.5$\AA$ while the distance between the Cu2-O2 and Bi-O4 planes increases by $\sim$0.3$\AA$ as compared to the experimental values. 
The Sr atoms keep their distance to the Cu2-O2 planes while the O3 atoms move away from the Cu2 ones towards the bismuth atoms. 

\begin{table}[hbt!]
\caption{\label{tabla1} Optimized and relaxed structural parameters of the tetragonal $Bi_2Sr_2Ca_2Cu_3O_{10}$. The experimental values were taken from reference~\cite{G}.}
\begin{tabular*}{0.231\textwidth}{ccc}\hline\hline
\multicolumn{3}{c}{Space group $I4/mmm$, z=2} \\
            &    Expt.      &  This work    \\\hline
{\em a}     & $3.823\AA$    & $3.843\AA$    \\
{\em c}     & $37.074\AA$   & $36.686\AA$   \\
{\em c/a}   & 9.70          & 9.55          \\\hline
atom        &      z        &     z          \\\hline
Bi          &   0.2109      &  0.2072        \\
Sr          &   0.3557      &  0.3682        \\
Ca          &   0.4553      &  0.4573        \\
Cu1         &   0.0000      &  0.0000        \\
Cu2         &   0.0976      &  0.0839        \\
O1          &   0.0000      &  0.0000        \\
O2          &   0.0964      &  0.0847        \\
O3          &   0.1454      &  0.1519        \\
O4          &   0.2890      &  0.2936        \\
\hline\hline
\end{tabular*}
\end{table}

The main difference between the crystal structure of Bi-2212~\cite{12b} and Bi-2223 is the presence of the new CuO$_2$ plane in the last one labeled in this work as Cu1-O1. The 
lattice parameter {\em a} is identical in both structures but the {\em c} one differs by $\sim7 \AA$. The presence of the new plane changes the inter-layer distance between 
neighboring cooper-oxygen planes. In Bi-2223, the inter-layer CuO$_2$ distance is $\sim0.38 \AA$ larger than in Bi-2212.
We will show the contribution in the electronic structure due Cu1-O1 plane.

\section{Results and discussion}
\subsection{Density of states}
\begin{figure}[hbt!]
\includegraphics[width=0.45\textwidth]{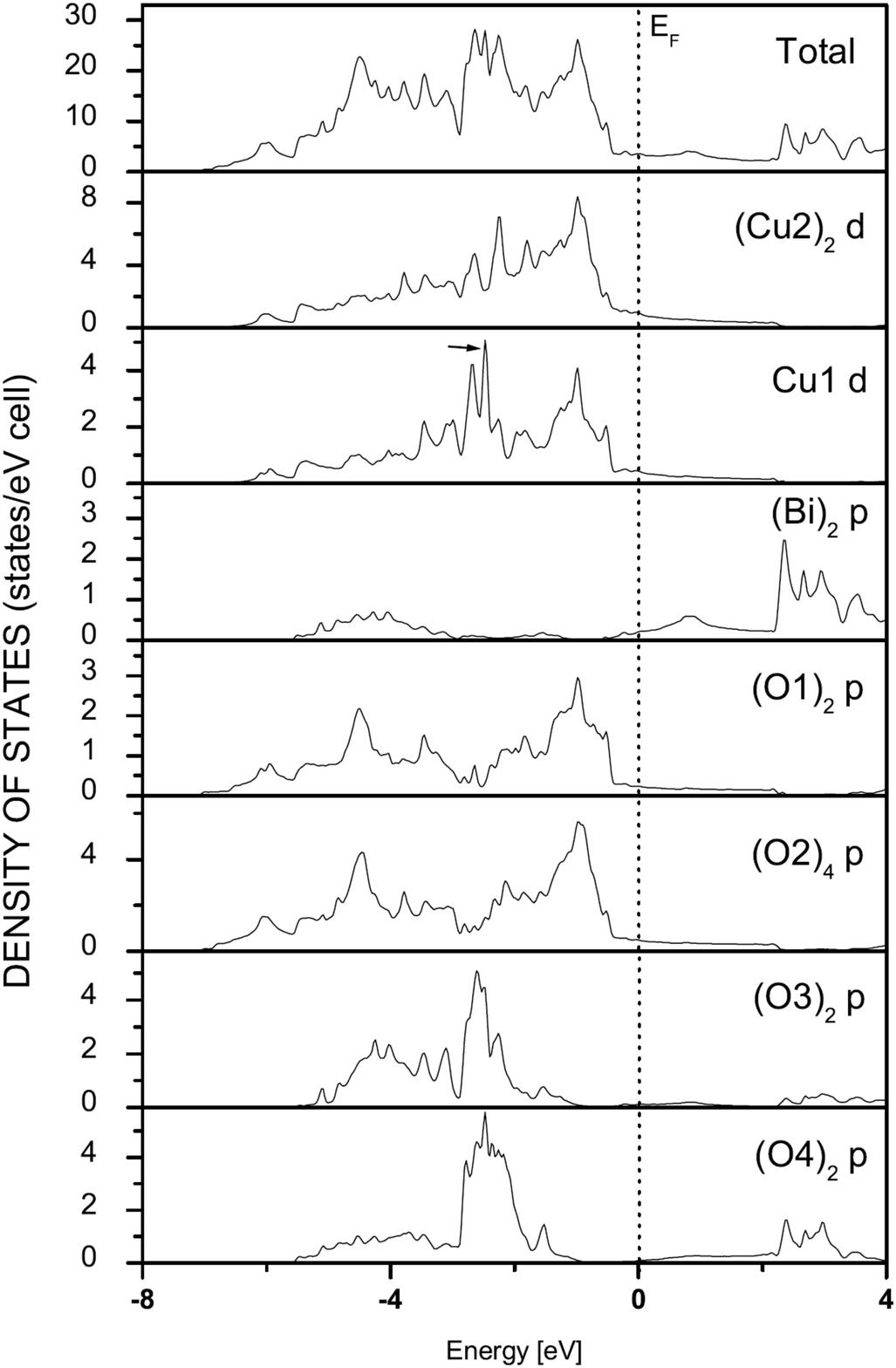} 
\caption{Total and atom-projected density of states for $Bi_2Sr_2Ca_2Cu_3O_{10}$. Note the change of scale for each atom contribution. The arrow in the pDOS of Cu1d indicates the 
peak at 2.49 eV below $E_F$.}
\label{Fig.1}
\end{figure}

Fig.~\ref{Fig.1} shows both the total Density of States (DOS) and the atom-projected densities of states (pDOS). We found that the Fermi level, $E_F$, falls in a region of low DOS. 
This behavior is similar to others Cu-O-based superconductors~\cite{12aa,13a,14a,15a}. As might be expected, the Bi-2223 compound maintains the same behavior as the DOS in the 
Bi-2212 compound~\cite{12b,13b,14b,15b}, the only difference is the contribution of new Cu1-O1 plane.
The total density of states at $E_F$, N($E_F$), for Bi-2223 is 3.55 states/(eV cell) which is larger than the one reported for Bi-2212 (2.1-3.3 states/(eV cell))~\cite{12b,14b} 
(see  Table\ref{tabla1a}). In the Bi-2223 compound a large contribution comes from the Cu2-O2, Cu1-O1 and Bi-O4 planes. The Bi-O plane contribution creates small electron pockets 
in the Fermi surface providing conduction electrons. These pockets, nevertheless, do not appear in the experimental results~\cite{pockets}.
It is important to note that the new Cu1-O1 plane contributes significantly to N($E_F$). On the other hand, comparing the atomic contributions to N($(E_F$) in both compound 
(see Table\ref{tabla1a}), we observe a larger contribution of Cu2-O2 planes in Bi-2223 and a similar one from the Bi-O4 planes.
The composition at $E_F$ of the Bi2223 DOS is mainly as follows, Cu2 $d_{x^2-y^2}$, O2 $p_{x,y}$, Cu1 $d_{x^2-y^2}$, O1 $p_{x,y}$,  Bi $p_{x,y}$, O4 $p_{x,y}$ and O3 $p_{x,y,z}$
states.

\begin{table}[hbt!]
\caption{\label{tabla1a} Atomic contributions to the density of states at the Fermi level, N($E_F$), for both Bi-2212 and Bi-2223. The values are given in units of states/eV-atom. 
The total N($E_F$) is in units of states/eV cell. The data for Bi-2212 were taken from Ref.~\cite{15b}. }
\begin{tabular*}{0.6\textwidth}{@{\extracolsep{\fill}}ccccccccc}\hline\hline
& \multicolumn{7}{c}{Atomic state}&  Total  \\
\cline{2-8}
Compound & Cu2 $d$ & Cu1 $d$& Bi $p$ & O1 $p$ & O2 $p$& O3 $p$ & O4 $p$& N($E_F$)\\
\cline{1-1} \cline{2-2} \cline{3-3} \cline{4-4} \cline{5-5} \cline{6-6} \cline{7-7} \cline{8-8} \cline{9-9}
Bi-2212  & -    & 0.33      & 0.17   & 0.16  & 0.07   & 0.07&- & 2.88     \\
Bi-2223  & 0.47    & 0.44   & 0.10   & 0.12   & 0.12  & 0.06   & 0.04& 3.55  \\
\hline\hline
\end{tabular*}
\end{table}

In Fig.~\ref{Fig.1} we observe that the contribution of the d-states to the pDOS from the Cu1 and Cu2 atoms is very similar. The only difference is the peak at 2.49 eV below $E_F$ 
which corresponds to the contribution of the Cu1 $d_{xz,yz}$ states. The occupied bandwidth of the Cu-O planes is $\sim$7 eV. Also observe the wide bandwidth of $\sim$9 eV which 
is typical of the $dp\sigma$ (bonding and anti-bonding) bands from the two-dimensional CuO$_2$ layers~\cite{14a}. This behavior is similar in the Bi-2212 
compound~\cite{12b,13b,14b,15b}.
The important contribution to the DOS above  $E_F$  comes mainly from the Bi $p$ states with a contribution of the O3 $p$ and O4 $p$ states. This oxygen states have a major 
contribution below at $E_F$. The DOS has a minor contribution from both Sr and Ca atoms due to their strong ionic character (Not shown in the Fig.~\ref{Fig.1} ).

We also calculated the total magnetic moment per cell for Bi-2223 compound and obtained 0.011 $\mu_B$, whose main contribution is due to the Cu atoms. This implies that the 
compound does not  exhibit a significant magnetic character at $T=0K$. This fact has implications for the explanation of the re-entrant behavior mentioned above.

\subsection{Band structure}
\begin{figure}[ht]
\includegraphics[width=1\textwidth]{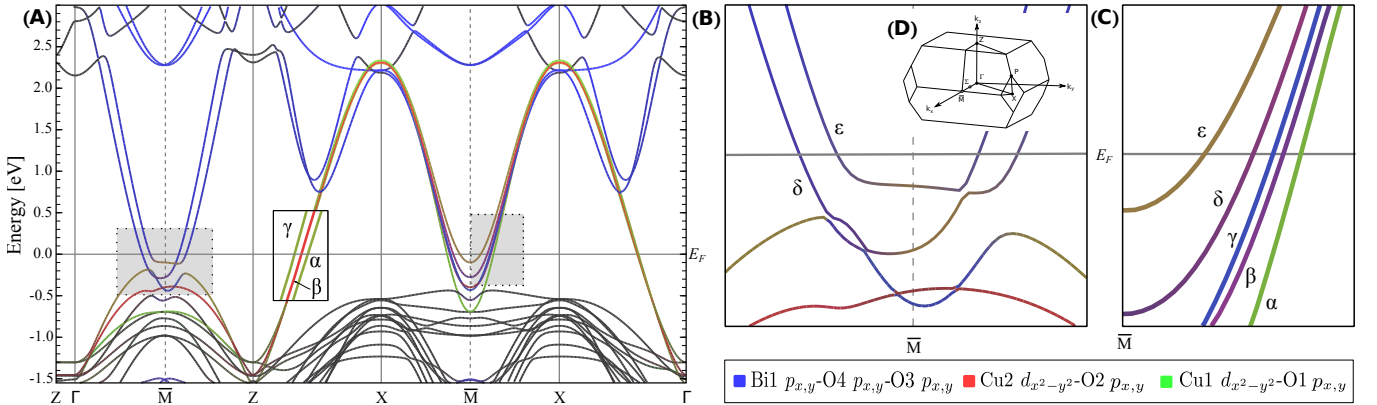}
\caption{(Color online) The band structure of the $Bi_2Sr_2Ca_2Cu_3O_{10}$ compound in the extended zone scheme. Note that the behavior of the bands are similar along the X-Z
and the X-$\Gamma$ directions. The shaded areas in the figure (A) are amplified in Figures (B) and (C) respectively. In (D) we show the first brillouin zone for a 
body-centered tetragonal structure for completeness.}
\label{Fig.2} 
\end{figure}

The band structure of the Bi-2223 compound is shown in the extended zone scheme in Fig.\ref{Fig.2}. (Notice that the $\overline{M}$ point is the midpoint between the $\Gamma$ 
and Z points along the $\Sigma$ direction). This band structure has many features in common with the Bi-2212 compound~\cite{12b,13b,14b,15b}. In the band structure, the states 
just below  $E_F$ are primarily Cu2(3$d$),O2(2$p$),Cu1(3$d$) and O1(2$p$) states, with a small contribution from Bi(6$p$), O4(2$p$) and O3(2$p$) states. Above $E_F$, most of 
the states are Bi(6$p$), O4(2$p$) and O3(2$p$) with a minor contribution from states from the Cu-O planes. As it can be seen in the Fig.\ref{Fig.2}(A), the band dispersion 
in the $\Gamma$-Z direction (perpendicular to the basal plane) is minimal which means that the bands are strongly two dimensional.

There are five bands crossing at $E_F$ which are composed primarily of Cu2 $d_{x^2-y^2}$ and O2 $p_{x,y}$ (in red), Cu1 $d_{x^2-y^2}$ and O1 $p_{x,y}$ (in green), Bi $p_{x,y}$, 
O4 $p_{x,y}$ and O3 $p_{x,y}$ (in blue) states in Fig.\ref{Fig.2} (Color online). The hybridized states from these bands are represented by their respective color mixture and 
the black line represents the other states. In Table~\ref{Tabla2} we present in detail the contribution at $E_F$ from the different atomic states.

\begin{figure}[htb!]
     \includegraphics[width=0.35\textwidth]{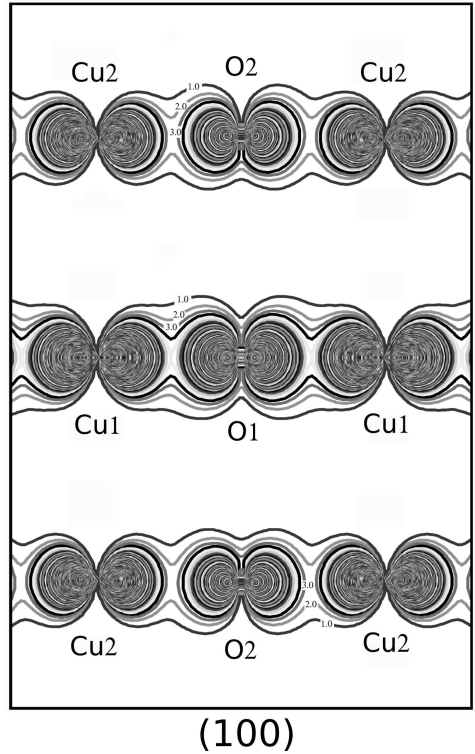} 
     \caption{\label{Fig.3} Charge density contour plots for the $\alpha$ band at $E_F$ in the  Z-X direction, on the (100) plane cutting the Cu1-O1 and Cu2-O2 bonds. Contours are 
     given on a linear scale which values are $10^{-3}e/a.u^3$.} 
\end{figure}

These bands cross $E_F$ in two regions, the first one between the Z-X(Y-$\Gamma$) points and the second one near the $\overline{M}$ point. In the first region (inset inside 
Fig.\ref{Fig.2}(A)) there are three nearly degenerate bands, two of them labeled as $\alpha$ and $\gamma$, are composed of Cu1 $d$, O1 $p$, Cu2 $d$, and O2 $p$ states (see 
Table~\ref{Tabla2}). In order to show the character of the $\alpha$ band at $E_F$, we plot in Fig.~\ref{Fig.3} the corresponding contour plots of the charge density, 
calculated at $k=2\pi(0.19/a, 0.19/b, 0.61/c)$ on the (100) plane cutting the Cu1-O1 and Cu2-O2 bonds. There we see that the Cu-O states are bonding very similar to the 
corresponding Cu-O states of the Bi-2212 compound~\cite{14b}. The other band (labeled as $\beta$) is also composed of bonding Cu2 $d$ and O2 $p$ states. These three bands 
(see inset inside Fig.~\ref{Fig.2}(A)) give the higher contribution to the DOS at the Fermi level. 

\begin{table*}[hbt!]
\caption{\label{Tabla2} Detailed contribution from the different atomic states to the bands at $E_F.$} 
\begin{tabular*}{0.95\textwidth}{@{\extracolsep{\fill}}cc | cccccccc}\hline\hline
                                                     &         &    Bi        & O4            &    O3       &  Cu1          & O1        &  \multicolumn{2}{c}{Cu2}  &  O2         \\
                                                                \cline{3-3}    \cline{4-4}      \cline{5-5}    \cline{6-6}    \cline{7-7}  \cline{8-9}        \cline{10-10}            
Direction                                            & Band    &  $p_{x,y}$   &  $p_{x,y}$    & $p_{x,y,z}$ & $d_{x^2-y^2}$ &  $p_{x,y}$& $d_{x^2-y^2}$  &$d_{z^2}$ & $p_{x,y}$   \\\hline          
\multirow{3}{*}{Z-X}                                 & $\alpha$ &      -       &     -         &     -       &       29$\%$  &   25$\%$  &     29$\%$     &    -     &   17$\%$    \\                    
                                                     & $\beta$ &      -       &     -         &     -       &       -       &     -     &     62$\%$     &    -     &   38$\%$    \\                  
                                                     & $\gamma$ &      -       &     -         &     -       &       28$\%$  &   24$\%$  &     29$\%$     &    -     &   19$\%$    \\\hline                    
$\Gamma$-Z ($\Sigma$)                                & $\delta$ &  44$\%$      &  24$\%$       &    25$\%$   &       -       &     -     &      4$\%$     &    -     &    3$\%$    \\\hline                    
$\Gamma$-$\overline{\text{M}}$                       & $\varepsilon$ &  33$\%$      &  9$\%$        &    19$\%$   &     6$\%$     &   3$\%$   &     16$\%$     &  6$\%$   &    8$\%$    \\\hline
$\overline{\text{M}}$-Z                              & $\varepsilon$&  47$\%$      &  13$\%$       &    27$\%$   &       -       &     -     &       6$\%$    &    -     &   11$\%$    \\\hline                                                                       
\multirow{5}{*}{$\overline{\text{M}}$-Y(X)}          & $\alpha$ &      -       &     -         &     -       &     44$\%$    &   16$\%$  &      29$\%$    &    -     &   11$\%$    \\                    
                                                     & $\beta$ &  18$\%$      &  12$\%$       &    8$\%$    &       -       &     -     &      45$\%$    &    -     &   17$\%$    \\                  
                                                     & $\gamma$ &  42$\%$      &  25$\%$       &    19$\%$   &     10$\%$    &    4$\%$  &       -        &    -     &    -        \\                  
                                                     & $\delta$ &  17$\%$      &  7$\%$        &    11$\%$   &       -       &     -     &      47$\%$    &    -     &   18$\%$    \\                  
                                                     & $\varepsilon$ &      -       &     -         &     -       &     24$\%$    &   8$\%$   &      46$\%$    &   7$\%$  &   15$\%$    \\\hline\hline                 
\end{tabular*}
\end{table*}

\begin{figure}[htb!]
     \includegraphics[width=0.35\textwidth]{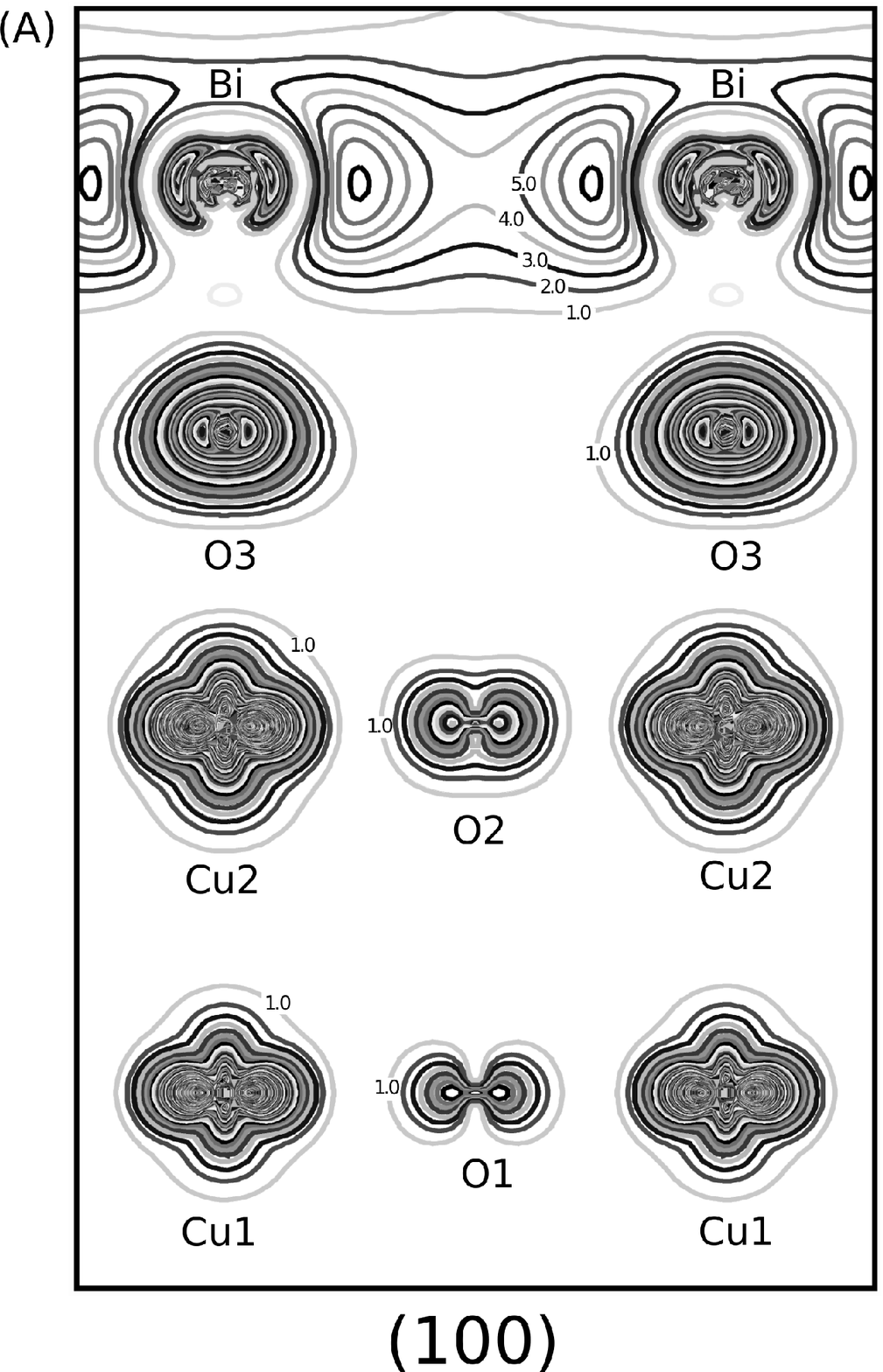} \\
      \medskip
     \includegraphics[width=0.35\textwidth]{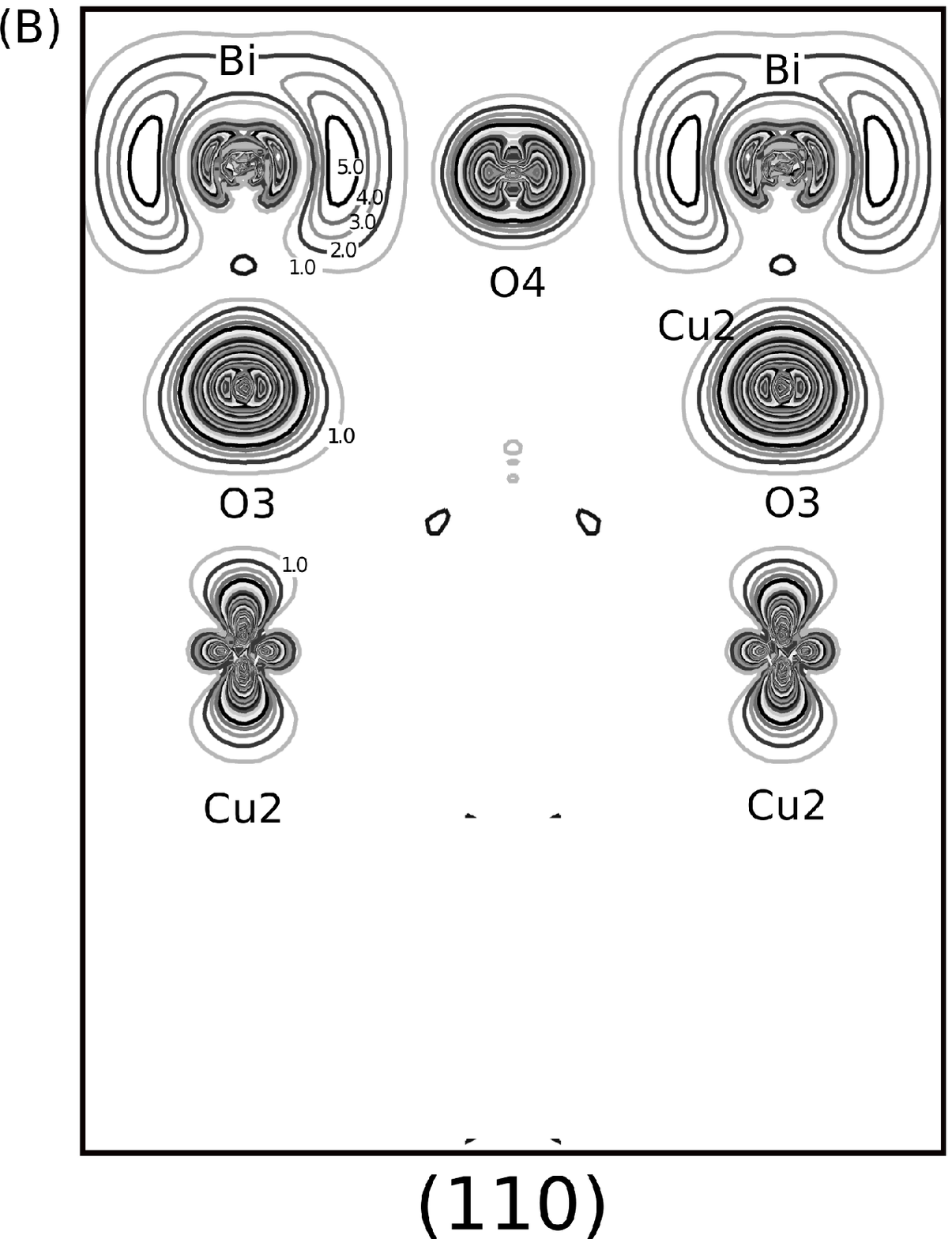} 
     \caption{\label{Fig.4} Charge density contour plots for the $\varepsilon$ band at $E_F$ in $\Gamma$-$\overline{\text{M}}$ direction, in the (100) and (110) planes. Note the 
     bonding character of Bi-Bi bonds and the antibonding character of Cu-O bonds in (100) plane. In (110) plane the contribution of the Cu1-O1 plane is not observed (see text). 
     Contours are given as in Fig.~\ref{Fig.3}.}
\end{figure}

In the second region (near the $\overline{M}$ point), the behavior of the bands are different along the $\Gamma$-Z and the X-Y directions (see shaded areas in 
Fig.~\ref{Fig.2}(A)). Further, as seen in Fig.~\ref{Fig.2}(B) there are two bands, labeled as $\delta$ and $\varepsilon$, crossing at $E_F$ in the $\Sigma$ direction, 
which are non symmetric around the $\overline{M}$ point. Between the $\Gamma$ and Z points, the $\delta$ band is composed of weakly bonding Bi $p$-O4 $p$ and anti-bonding 
Bi $p$-O3 $p$ states.
The $\varepsilon$ band between the $\Gamma$ and $\overline{M}$ points, is formed of anti-bonding Bi $p$-O4 $p$ and Bi $p$-O3 $p$ states and hybridizes with Cu2 $d$-O2 $p$ 
and Cu1 $d$-O1 $p$ states, with a small contribution from Cu2 $d_{z^2}$ states, while between the $\overline{M}$ and Z points it is composed of anti-bonding Bi $p$-O4 $p$ 
states and hybridizes with Cu2 $d$-O2 $p$ and O3 $p$ states.
In Fig.~\ref{Fig.4} we show the contour plots of the charge density in the (100) and (110) planes, cutting the Bi-O and Cu-O bonds corresponding to the band $\varepsilon$ 
in Fig.~\ref{Fig.2}(B) at $k=2\pi(0.42/a, 0, 0)$, very close to $E_F$. Fig.~\ref{Fig.4}(A) shows the strong Bi-Bi bonding character ($pp\sigma$) and the anti-bonding character 
of the Cu-O planes. 
In Fig.~\ref{Fig.4}(B) we show the contour plot of the charge density on the (110) plane. There we can see the anti-bonding character of the Bi-O and Cu2-O3 states. This 
behavior is similar to the one reported for Bi-2212 by Massidda et al.~\cite{14b}.

Around the $\overline{M}$ point but in the X-Z direction (see Fig.\ref{Fig.2}(C)), the $\alpha$, $\beta$, $\gamma$, $\delta$ and $\varepsilon$ bands have a different character. The 
band $\alpha$ is a combination of bonding Cu1 $d$-O1 $p$ and Cu2 $d$-O2 $p$ states (see Table\ref{Tabla2}). The bands $\beta$ and $\delta$ are composed of hybrids of bonding Cu2 
$d$-O2 $p$, antibonding Bi $p$-O4 $p$ and Bi $p$-O3 $p$ states. The band $\gamma$ is composed of weakly bonding Bi $p$-O4 and Bi $p$-O3 $p$ states, and hybridizes with antibonding 
Cu1 $d$-O1 $p$ states. Finaly, the band $\varepsilon$ has a similar behavior as the band $\alpha$ but with a small contributions of the Cu2 $d_{z^2}$ state at $E_F$ (see 
Table~\ref{Tabla2}).

In the Fig.\ref{Fig.2}(A), the copper-oxygen bands that cross at $E_F$ extend from 1.5 eV below to 2.3 eV above $E_F$. These bands have their maximum energy at the X point, and 
their minimal energy at the Z point and are anti-bonding. These bands present a strong Cu-O $dp\sigma$ character. Around the $\overline{M}$ point, the Bi-O bands extend to about 
0.57 eV below $E_F$ and hybridize with Cu2 $d_{x^2-y^2}$-O2 $p_y$ states. 
The Bi-O bands presents a $pp\sigma$ character. The contribution of the Bi-O states at $E_F$ are due to the interaction with the O3 and the Cu2-O2 planes. The Bismuth bands present an 
interesting feature around $\Gamma$ and Z. Exactly at those points, the band derives from only $p_z$ state, while around those points the band derives from combinations of all 
$p$ states.

The bands around the X(Y) point are $\sim$0.54 eV below $E_F$ while in the Bi-2212 compound this bands lie about $\sim$0.1 eV below $E_F$~\cite{12b,13b,14b,15b}. These bands 
are primarily antibonding Cu1 $d_{xz,yz}$-O1 $p_{z}$ and Cu2 $d_{xz,yz}$-O2 $p_{z}$ states with a $pp\pi$ character. Taking the band structures of $Bi_2Sr_2Ca_{n-1}Cu_nO_y$ 
compounds with $n=1$ and $2$, calculated by Sterne and Wang~\cite{16b} and comparing with our calculation ($n=3$), we observed that the number of bands crossing at $E_F$ in 
$\Gamma$-X direction is proportional to the number of Cu-O planes. A similar idea had been outlined by Mori et al.~\cite{mori}. We also noted that the energy below $E_F$ of 
the Bi-O bands around $\overline{M}$ point is depper proportionally to the number of Cu-O planes.

\subsection{Fermi surface}

\begin{figure}[htb!]
     \includegraphics[width=0.5\textwidth]{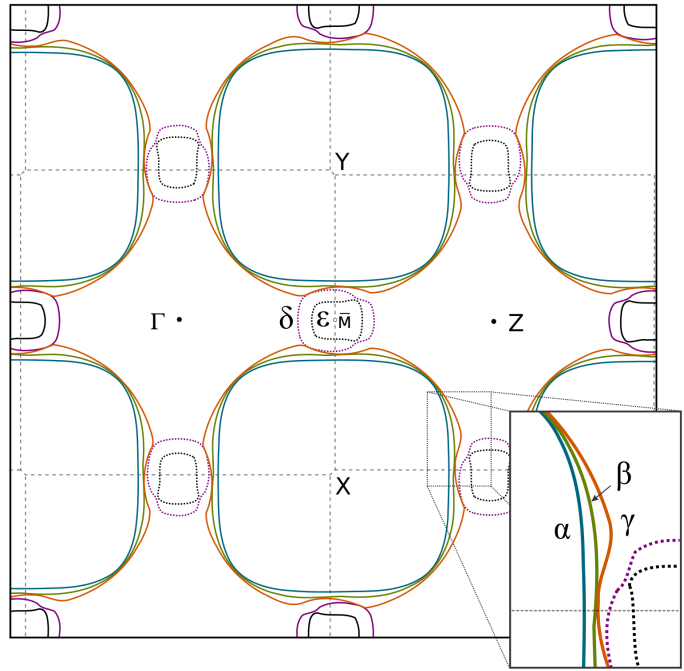} 
    \caption{\label{Fig.5} (Color online) Fermi Surface at $k_z=0$ of $Bi_2Sr_2Ca_2Cu_3O_{10}$ in an extended zone scheme. The Bi-O pockets are represented by violet and 
    black lines (dashed lines).}
\end{figure}

In Fig.\ref{Fig.5} we show the Fermi Surface (FS) of the Bi-2223 compound in an extended zone scheme. This FS has in general the same behavior to the one reported for the Bi-2212 
compound, with the similar highly anisotropic low-dimensionality~\cite{14b,15b,clau}. Our calculation presents two additional surfaces. 
Around the X(Y) there are three not quite degenerate hole surfaces labeled as $\alpha$, $\beta$ and $\gamma$ (corresponding to the respective bands in Fig.\ref{Fig.2}). 
In $\Gamma$-X and equivalent directions, the $\alpha$ and $\gamma$ surfaces consists of hybrids of Cu1-O1 and Cu2-O2 states and the $\beta$ surface is composed of Cu2-O2 
states. The $\alpha$ and the $\beta$ surfaces are close to be rounded squares. As we approach from the X(Y) to the $\overline{\text{M}}$ point, the $\beta$ and the $\gamma$ 
surfaces get an additional contribution from Bi-O4 states (see Table~\ref{Tabla2}). As we can see from Fig.\ref{Fig.5} the $\alpha$ surface has more nesting than $\beta$ 
and $\gamma$ surfaces. 

\begin{figure}[htb!]
     \includegraphics[width=0.5\textwidth]{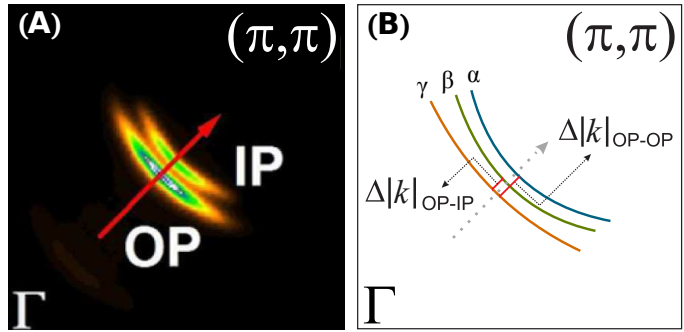} 
    \caption{\label{Fig.FS} (Color online) (A) The Fermi surface (FS) of Bi-2223 calculated by ARPES~\cite{M} in the nodal direction. (B) Schemetic FS showing the band splitting and
    the $\Delta |k|$ calculated in this work.}
\end{figure}

The FS of Bi-2223 calculated by ARPES is shown in ref.~\cite{M}. In that work they found two surfaces on the $\Gamma$-X(Y) direction (see Fig.\ref{Fig.FS}), that they call 
outher copper planes (OP) and inner copper plane (IP) and suggest the possibility that OP's are degenerate. The full width at half maximum (FWHM) of the momentum distribution 
curve (ARPES resolution) for the OP is $\sim0.011 \AA$ and the IP is $\sim0.0074 \AA$, at $E_F$. Other experimental work with the same technique~\cite{K,L} does not report this 
band splitting. 

We identified these IP and OP as $\alpha$ and $\beta$, $\gamma$ surfaces respectively in our FS. On $\Gamma$-X direction we calculated the differences in momentum 
$\Delta |k|_{OP-IP}$ and $\Delta |k|_{OP-OP}$(see Fig.\ref{Fig.FS}) an found $\sim0.005 \AA$ and $\sim0.01 \AA$ respectively. Comparing these differences with the ARPES 
resolution in the work just mentioned, it is clear that it cannot resolve the existence of the three bands separately. Our results support the idea that the IP is composed of Cu2-O2 states, and 
the OP's are composed of hybrids from Cu1-O1 and Cu2-O2 states.

Now, around the $\overline{\text{M}}$ point we observe two surfaces (Bi-O pockets) in comparation with the single surface observed in the Bi-2212 compound. These surfaces, labeled 
as $\delta$ and $\varepsilon$ (see Fig.~\ref{Fig.5}) form small closed electron surfaces. The $\delta$ surface is close to be a rounded square with a small convexity pointing towards Z 
point, while $\varepsilon$ surface is almost a rounded rectangle. In the experimental reports the Bi-O planes show always a nonmetallic character~\cite{pockets}. In the 
theoretical calculations this is called the ``Bi-O pocket problem''. 

Finaly, around $\Gamma$ and Z points there are electron surfaces that look like a between them, this characteristic is due to highly two-dimensionality of the system, 
however from the band structure and the FS of the Bi-2223 compound we observe less two dimensional behavior than the reported for the Bi-2212 compound~\cite{14b,15b}.

\section{Conclusions}

We presented here a detailed analysis of the electronic properties of the tetragonal ($I4/mmm$) Bi-2223 compound. To the best of our knowledge there is no previous theoretical 
calculation for this compound. Our calculation was done using the Full-potential linearized augmented plane wave method plus Local orbitals within the Local density approximation 
using Wien2k code.

We studied the contribution of the Cu1-O1 plane to the electronic properties of the Bi-2223 compound. This plane has an important contribution to the DOS at the $E_F$ 
(0.56 states/eV-atom) which is similar to the one of the Cu2-O2 plane. Compared to Bi-2212~\cite{15b}, Bi-2223 presents a higher DOS at $E_F$. This is due to the new Cu-O plane.

Our calculated band structure present Bi-O bands at the $E_F$. This so called Bi-O pocket problem is in desagreement with the experimental results. This problem also appears in the 
theoretical calculations concerning Bi-2212 and Bi-2201. 
Taking the band structures of Bi-2212 and Bi-2201~\cite{16b} and comparing with our calculation (Bi-2223), we observed that the number of bands crossing at $E_F$ in 
$\Gamma$-X direction is proportional to the number of Cu-O planes. This is in agreement with the idea outlined by Mori et al.~\cite{mori}. We also noted that the energy below 
$E_F$ of the Bi-O bands around $\overline{M}$ point is depper proportionally to the number of Cu-O planes. 

The Fermi surface (FS) calculated in this work presents a good agreement with the experimental result calculated, in the nodal direction, by angle-resolved photo-emission 
spectroscopy (ARPES)~\cite{M}.  

Our work is useful per se and contributes to the understanding of the re-entrant behavior found in the ceramic composite of $CdS/Bi_2Sr_2Ca_2Cu_3O_{10}$.

\section{Acknowledgments}
The authors acknowledge to the GENERAL COORDINATION OF INFORMATION AND COMMUNICATIONS TECHNOLOGIES (CGSTIC) at CINVESTAV for providing HPC resources on the Hybrid Cluster 
Supercomputer "Xiuhcoatl", that have contributed to the research results reported within this paper.

\end{document}